\documentclass[ reprint, 
 aps,
 pra,
 showkeys,
 nofootinbib
]{revtex4-1}
\usepackage{amsfonts}
\usepackage{amssymb}
\usepackage{amsmath}
\usepackage{graphicx}
\usepackage{subfigure}
\usepackage{dcolumn}
\usepackage{bm}
\usepackage{setspace}
\usepackage[hang]{footmisc}
\usepackage[sort&compress]{natbib}
\usepackage{multirow}
\usepackage[sort&compress]{cleveref}
\usepackage{nomencl}
\usepackage[english]{babel}
\usepackage{xcolor, soul}
\usepackage{marginnote}

\sethlcolor{yellow}

\crefname{equation}{\unskip}{\unskip}
\crefname{figure}{\unskip}{\unskip}
\crefname{section}{\unskip}{\unskip}
\crefname{subsection}{\unskip}{\unskip}
\begin{document}
\let\ref\cref

\title{Elastohydrodynamics for soft solids with surface roughness: transient effects}

\author{M. Scaraggi}
\thanks{michele.scaraggi@unisalento.it}
\affiliation{DII, Universita del Salento, 73100 Monteroni-Lecce, Italy, EU}
\affiliation{PGI-1, FZ J\"ulich, Germany, EU}
\author{L. Dorogin}
\affiliation{PGI-1, FZ J\"ulich, Germany, EU}
\author{J. Angerhausen}
\affiliation{IFAS, RWTH University, Germany, EU}
\author{H. Murrenhoff}
\affiliation{IFAS, RWTH University, Germany, EU}
\author{B.N.J. Persson}
\thanks{b.persson@fz-juelich.de}
\affiliation{PGI-1, FZ J\"ulich, Germany, EU}
\affiliation{www.MultiscaleConsulting.com}

\begin{abstract}
A huge number of technological and biological systems involves the lubricated contact between rough surfaces of soft solids in relative 
accelerated motion. Examples include dynamical rubber seals and the human
joints. In this study we consider an elastic cylinder with random surface roughness
in accelerated sliding motion on a rigid, perfectly flat (no roughness)
substrate in a fluid. We calculate the surface deformations, interface separation and the contributions to the
friction force and the normal force from the area of real contact and from the fluid. The driving velocity profile as a
function of time is assumed to be either a sine-function, or a linear multi-ramp function. We show how the squeeze-in and squeeze-out
processes, occurring in accelerated sliding, quantitatively affect the Stribeck curve with respect to the steady sliding.
Finally, the theory results are compared to experimental data. 
\end{abstract}

\maketitle
\makenomenclature


{\bf 1 Introduction}

The nature of the lubricated contact between soft elastic bodies is one of the central topics in tribology\cite{Persson0,Meyer}, with
applications to the human joints and eyes\cite{Greg}, dynamic rubber seals, and the tire-road interaction, to name just a few examples. 
However, these problems are also very complex involving large elastic deformations and fluid flow between narrowly spaced walls 
and in irregular channels\cite{Mueser}. For smooth spherical or cylindrical bodies in steady sliding on flat lubricated
substrates (i.e. without surface roughness), such {\it elastohydrodynamic} problems are now well understood\cite{Dowson,elasto},
at least as long as interface energies are unimportant. 
However, for more common cases involving non-steady sliding, with surfaces with roughness on many 
length scales, and with non-Newtonian fluids, rather little is known\cite{PS1}.  

In a series of papers two of us have shown how one may take into account the surface roughness when studying
the influence of a fluid on the sliding (constant velocity) of 
an elastic cylinder (or sphere), against another solid with a nominally flat surface\cite{PS0,PS1,PS2,PS3}.
Using the same approach we have also studied the fluid squeeze-out between elastic solids\cite{SP0,SP1}.
In this paper we study the more general case of accelerated sliding motion. In particular,
here we investigate the contact between a lubricated stationary elastic cylinder and a rigid nominally flat substrate in accelerated motion. 
We calculate the surface deformations, interface separation and the contributions to the
friction force and the normal force from the area of real contact and from the fluid. The driving velocity profile as a
function of time is assumed to be either a sine function, or a linear multi-ramp function. We also calculate the steady state
friction coefficient as a function of sliding speed (the Stribeck curve), and we compare it with the friction resulting from the
accelerated motion, the latter affected by the squeeze-in and squeeze-out dynamics. In all cases we assume the surface roughness
to show a self-affine fractal content, whereas the fluid is treated as a Newtonian fluid i.e. the fluid viscosity
is assumed independent of the shear rate in the present study.

The manuscript is outlined as follows. In Sec. 2 we summarize the mean field lubrication model.
In Sec. 3 we show theory results of the sliding kinematics for a sinus motion and a linear multi-ramp motion, and in particular
we shed light on the squeeze-in and -out dynamics effects on the friction. 
In Sec. 4  we compare the theory predictions with experimental results. Sec. 5 contains the summary and conclusions.

\begin{figure}
\includegraphics[width=0.8\columnwidth]{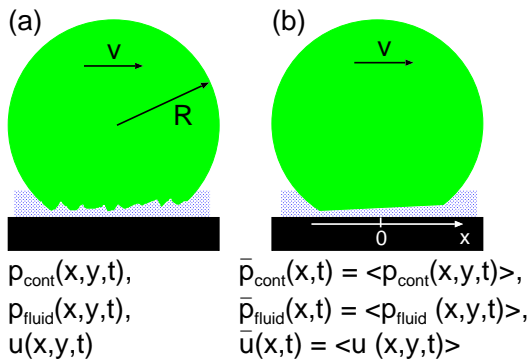}
\caption{\label{twoBALL.eps}
(a) Schematic of a rubber ball with a rough surface sliding on a smooth rigid substrate surface.
Physical quantities like the contact pressure, the fluid pressure and the interface separation
varies rapidly in space over many decades in length scales due to the nature of the surface roughness.
The complex situation in (a) can be mapped on a simpler situation (b) where the fluid and contact pressures, and
the surface separation, are locally-averaged quantities, which varies slowly in space on the length scale of
the surface roughness. Those averaged quantities obey to modified fluid flow equations which 
contain two functions, denoted as flow factors,
which depend on the locally averaged surface separation, and which are mainly determined by the surface roughness.}
\end{figure}

\vskip 0.3cm
{\bf 2 Theory}

\vskip 0.1cm
{\bf 2.1 Equations of motion}

We consider the simplest problem of an elastic cylinder (length $L$ and radius $R$, with $L>>R$)
with a randomly rough surface sliding on a 
rigid solid with a smooth (no roughness) flat surface. We assume that the sliding occurs in the
direction perpendicular to the cylinder axis, and we introduce a coordinate system
with the $x$-axis along the sliding direction 
and with $x=0$ corresponding to the cylinder axis position, see the schematic of Fig. \ref{twoBALL.eps}. 
The cylinder is squeezed against the substrate
by the normal force $F_{\rm N}$, and at the position $x$ in the contact region between the cylinder and the substrate
occur a nominal (locally averaged) contact pressure (see Fig. \ref{twoBALL.eps}) 
$$p_0 (x,t)=\bar p_{\rm cont}(x,t) + \bar p_{\rm fluid}(x,t), \eqno(1)$$
where $\bar p_{\rm cont}$ is the pressure due to the direct solid-solid interaction and $\bar p_{\rm fluid}$ is the
fluid pressure. 
The bar indicates that both pressures have been averaged over the surface roughness, e.g., 
$\bar p_{\rm fluid} (x,t) = \langle p_{\rm fluid} (x,y,t) \rangle$.
We consider a constant normal load so that 
$$\int_{-\infty}^{\infty} d x \ p_0(x,t) = {F_{\rm N}\over L}.\eqno(2)$$ 
Let $\bar u(x,t)$ denote the (locally averaged) separation between the surfaces. For $\bar u > h_{\rm rms}$, where
$h_{\rm rms}$ is the root-mean-square (rms) roughness parameter, 
$$\bar p_{\rm cont}(x,t) \approx \beta E^* {\rm exp} \left ( -\alpha {\bar u(x,t) \over h_{\rm rms}}\right ),\eqno(3)$$
where $\alpha$ and $\beta$ are described in Ref. \cite{PS.intsep}.
Eq. (3) is valid for large enough $\bar u$. Since an infinite high pressure is necessary in order
to squeeze the solids into complete contact we must have $p_{\rm cont} \rightarrow \infty$ as $\bar u \rightarrow 0$. This is, of course,
not obeyed by (3), and in our calculations we therefore use the numerically calculated relation $p_{\rm cont} (\bar u)$
which reduces to (3) for large enough $\bar u$. 

The macroscopic gap equation is determined by simple geometrical considerations. Thus, assuming the cylinder
deformation to be within the Hertz regime for elastic solids, the gap equation reads
$$\bar u(x,t)=u_0(t)+{x^2\over 2R} -{2\over \pi E^*}\int_{-\infty}^{\infty} dx' \ p_0 (x',t) {\rm ln}\left |
{x-x'\over x'}\right |. \eqno(4)$$
In addition the pressure $p_0(x,t)$ must satisfy the total normal load conservation condition (2).

Finally, we need an equation which determines the fluid pressure $\bar p_{\rm fluid}(x,t)$. The fluid flow is usually determined by the
Navier Stokes equation, but in the present case of fluid flow in a narrow gap between the solid walls, the equation can be simplified
resulting in the so called Reynolds equation. For surfaces with roughness on many length scales, this equation is also inconveniently too complex, numerically, to be directly solved.
However, when there is a separation of length scales, i.e., when the longest (relevant) surface roughness wavelength component is much shorter
than the width (in the sliding direction) of the nominal cylinder-flat contact region, it is possible to eliminate the surface roughness and obtain a
modified (or effective) Reynolds equation describing the locally averaged fluid velocity and pressure fields. Such equations
are characterized by two correction factors, namely $\phi_{\rm p}$ 
(pressure flow factor) and $\phi_{\rm s}$ (shear flow factor), which are mainly determined by the surface roughness and depend on the locally averaged surface
separation $\bar u$. Thus, the effective 2D fluid flow current 
$${\bf J} =  -{\bar u^3 \phi_{\rm p}(\bar u) \over 12 \eta} \nabla \bar p_{\rm fluid} +{1\over 2} \bar u {\bf v} +{1\over 2} h_{\rm rms} \phi_{\rm s} (\bar u) {\bf v}\eqno(5)$$
satisfies the mass conservation equation
$${\partial \bar u \over \partial t} +\nabla \cdot {\bf J} = 0. \eqno(6)$$
Substituting (5) in (6), and writing ${\bf v} = v_0 \hat x$, gives the modified Reynolds equation:
$${\partial \bar u \over \partial t} = {\partial \over \partial x} \left [ {\bar u^3 \phi_{\rm p}(\bar u) \over 12 \eta}  {\partial \bar p_{\rm fluid} \over \partial x} 
-{1\over 2} \bar u v_0 -{1\over 2} h_{\rm rms} \phi_{\rm s} (\bar u) v_0\right ].\eqno(7)$$

The equations (1), (2), (3), (4), and (7) represent 5 equations for the 5 unknown variables $p_0$, $\bar p_{\rm cont}$,
$\bar p_{\rm fluid}$, $\bar u$ and $u_0$. We note that (7) is solved with Cauchy boundary conditions, whereas the
macroscopic cavitation is set by requiring\footnote{We note that for soft elastic solids, like rubber, we have shown in Ref. \cite{PS0} that including cavitation or not has no drastic effect on the result, and in
particular the friction coefficient as a function of the sliding speed, $\mu = \mu (v)$, is nearly unchanged.}
 $\bar p_{\rm fluid}\geq 0$.

A brief note on the numerical procedure. Eq. (7) is discretized in the time with Crank-Nicolson approach and automated stepping, whereas central differences and structured mesh are adopted for the spatial derivatives. The resulting non-linear system of equations is then linearized and numerically solved as for the generic steady lubricated contact described in \cite{PS1}.

\vskip 0.1cm
{\bf 2.2 Frictional shear stress and friction force}

The friction force acting on the bottom surface can be obtained by integration of
the frictional shear stress over the bottom surface. The frictional shear stress has a contribution
from the area of contact $\tau_{\rm cont} (x,y,t)$ and another from the fluid $\tau_{\rm fluid} (x,y,t)$.
Because of the multiscale surface roughness both quantities varies rapidly in space. However,
one can eliminate (integrate out) the roughness and obtain effective (locally averaged)
contact and fluid shear stresses so the total effective shear stress is
$$\bar \tau = \bar \tau_{\rm cont} + \bar \tau_{\rm fluid}.\eqno(8)$$
For the cylinder geometry we consider, $\bar \tau$, $\bar \tau_{\rm cont}$ and  $\bar \tau_{\rm fluid}$ are
independent of the $y$-coordinate, i.e., they depend only on $x$ and the time $t$. The contribution from the area of contact
$\bar \tau_{\rm cont} = - \tau_1 A(x,t)/A(0)$ depend on the relative contact area $A(x,t)/A_0$, which we calculate using the Persson
contact mechanics theory. For simplicity we assume below that the shear stress $\tau_1$  is independent of the sliding speed.

The frictional shear stress ${\tau}_{\mathrm{fluid}}$
originating from the fluid is given by
\begin{equation*}
\tau _{\mathrm{fluid}}=\eta {\frac{\partial v_{x}}{\partial z}}.\eqno(9)
\end{equation*}%
Using the lubrication approximation this gives\cite{PS1}:
\begin{equation*}
\tau _{\mathrm{fluid}}\left( \mathbf{x}\right) =-\frac{\eta 
\mathbf{v}_{0}}{u(\mathbf{x})}-{\frac{1}{2}}u(\mathbf{x})\nabla p(\mathbf{x}%
).\eqno(10)
\end{equation*}%
Averaging over the surface roughness results in an effective fluid shear stress
\begin{equation*}
\bar{\tau}_{\mathrm{fluid}}=-\left( \phi _{\mathrm{f}}+\phi
_{\mathrm{fs}}\right) {\frac{\eta _{0}\mathbf{v}_{0}}{\bar{u}}}-\frac{1}{2}%
\phi _{\mathrm{fp}}\bar{u}\nabla \bar p_{\rm fluid},\eqno(11)
\end{equation*}%
where the friction factors $\phi _{\mathrm{f}}$, $\phi _{\mathrm{fs}}$ and $\phi _{\mathrm{fp}}$
depend on the average interfacial separation $\bar u$. In Ref. \cite{PS1}
we derived expressions for $\phi _{\mathrm{f}}$, $\phi _{\mathrm{fs}}$ and $\phi _{\mathrm{fp}}$
which we use in the calculations presented below.

One particular important factor is
$$\phi_{\rm f} = {\bar u \over \eta_0} \left \langle {\eta \over u({\bf x})}\right \rangle, \eqno(12)$$
where $\eta_0$ is the low shear rate fluid viscosity, and where $\eta $ is the viscosity at the
shear rate $\dot \gamma$. It is very important to note that $\phi_{\rm f}$ can be very large, 
and can have a very strong influence on the friction force (see Sec. 3.1 below). Neglecting shear
thinning, it follows from (12) that when the separation $u({\bf x})$ is constant, $\phi_{\rm f}=1$.
The following arguments show, however, that if the fluid film thickness varies strongly with ${\bf x}$, 
which will always be the case when the sliding speed becomes so low that asperity 
contacts start to occur, $\phi_{\rm f}$ can be much larger than unity.

\begin{figure}
\includegraphics[width=0.6\columnwidth]{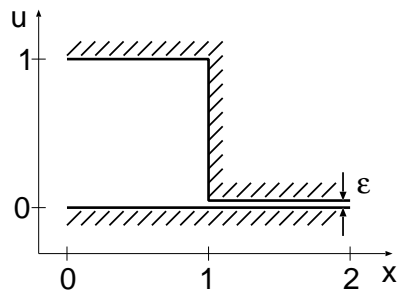}
\caption{\label{explain.eps}
The interfacial separation $u(x)$ as a function of the lateral coordinate $x$. 
We assume $\epsilon << 1$. Hence the average interfacial
separation $\bar u = \langle u \rangle = (1 + \epsilon)/2 \approx 1/2$, while the average of the
inverse of the separation is $\langle 1/u \rangle = (1+ 1/\epsilon )/2 \approx 1/(2 \epsilon) >> \bar u$. 
Hence in this case (assuming no shear thinning) $\phi_{\rm f} = \bar u \langle 1/u \rangle \approx 1/\epsilon >> 1$.
}
\end{figure}

In Fig. \ref{explain.eps} we illustrate the origin of why $\phi_{\rm f} >> 1$ in some cases.
Assume for simplicity no shear thinning so that $\phi_{\rm f} = \bar u \langle 1/u \rangle$.
Assume that the interfacial separation $u(x)$, as a function of the lateral coordinate $x$, takes the
form shown in Fig. \ref{explain.eps} with $\epsilon << 1$. Hence the average interfacial
separation $\bar u = \langle u \rangle = (1 + \epsilon)/2 \approx 1/2$, while the average of the
inverse of the separation is $\langle 1/u \rangle = (1+ 1/\epsilon )/2 \approx 1/(2 \epsilon) >> \bar u$. 
Hence in this case $\phi_{\rm f} = \bar u \langle 1/u \rangle \approx 1/\epsilon >> 1$.

\begin{figure}
\includegraphics[width=1.0\columnwidth]{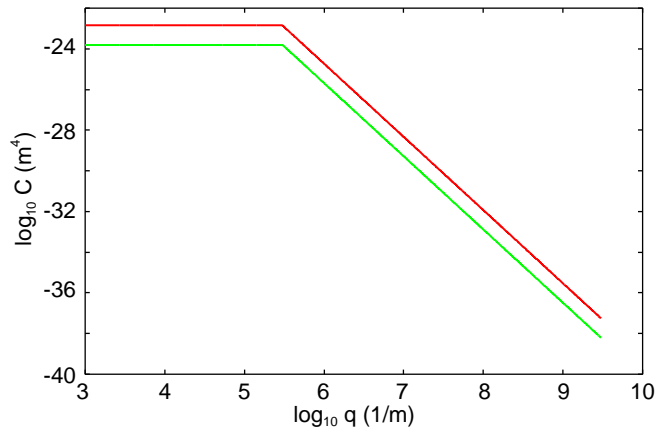}
\caption{\label{Powerspectrum_Roughness3micron_vs_Roughness1micron-Reference.eps}
Example of isotropic surface roughness power spectrum for surfaces: red curve for the surface with $h_{\rm rms}=3 \ {\rm \mu m}$, green curve for the one with $h_{\rm rms}=1  \ {\rm \mu m}$.
}
\end{figure}

\begin{figure}
\includegraphics[width=1.0\columnwidth]{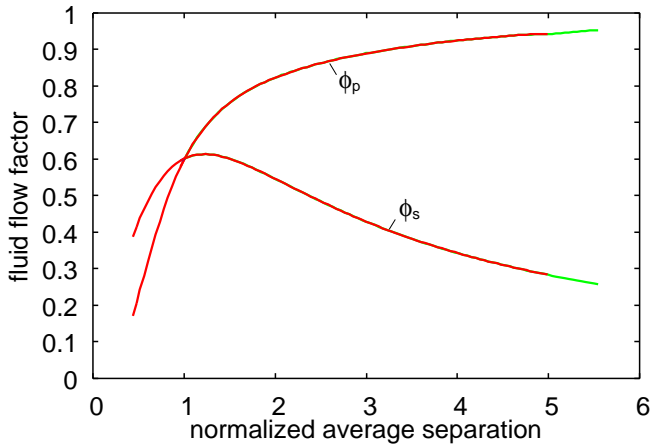}
\caption{\label{FluidFlowFactors_Roughness3micron_vs_Roughness1micron-Reference.eps}
Fluid pressure flow factor $\phi_{\rm p}$ and shear stress flow factor $\phi_{\rm s}$ as functions of average separation normalized by $h_{\rm rms}$.
(The green tails to be removed! or put the both curves in black.)
}
\end{figure}
\begin{figure}
\includegraphics[width=1.0\columnwidth]{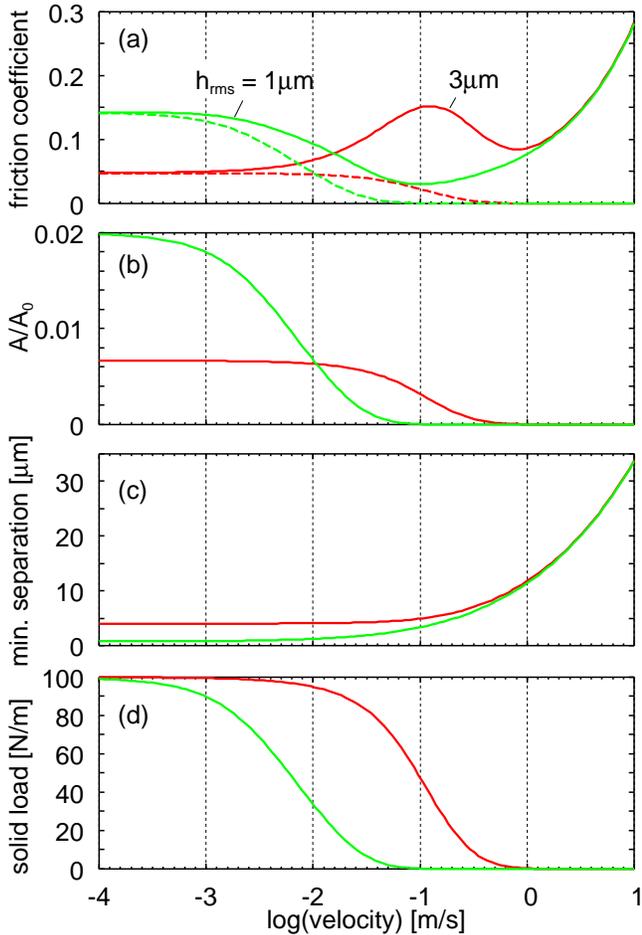}
\caption{\label{StribeckCurve_Roughness3micron_vs_Roughness1micron-Reference.eps}
The (a) friction coefficient, (b) actual area of contact, (c) minimum separation and (d) solid load, 
as functions of velocity for surfaces with different roughness: red curves for $h_{\rm rms}=3 \ {\rm \mu m}$
and green curves for $h_{\rm rms}=1 \ {\rm \mu m}$. Dashed lines show the contribution of solid-solid contact to the friction coefficient.
}
\end{figure}
\begin{figure}
\includegraphics[width=1.0\columnwidth]{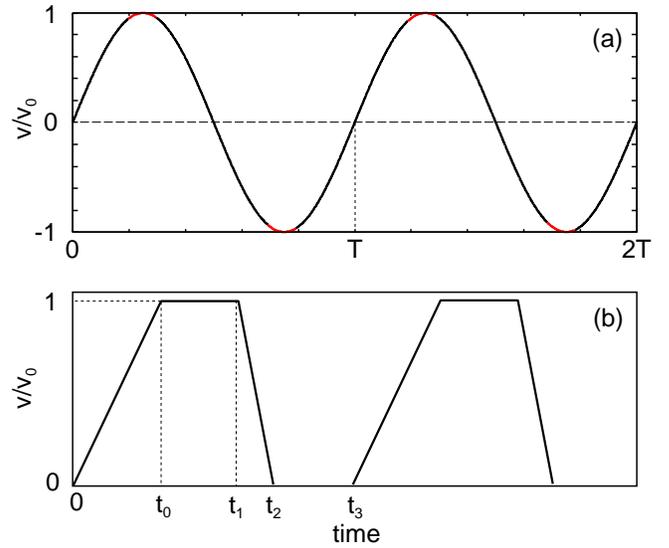}
\caption{\label{sinus.eps}
The sliding speed as a function  of time for (a) sinus time-dependency and for (b) linear multi-ramp.
}
\end{figure}

\vskip 0.3cm
{\bf 3 Numerical results}

We consider the sliding of an elastic cylinder (radius $R=4 \ {\rm mm}$) with a randomly rough surface on a rigid, perfectly smooth substrate. 
The cylinder has the Young's modulus $E=3 \ {\rm MPa}$ and Poisson ratio $\nu = 0.5$. We consider two cases where
the cylinder rms surface roughness amplitude is $h_{\rm rms}=1  \ {\rm \mu m}$ and $h_{\rm rms}=3 \ {\rm \mu m}$, respectively.
The surface roughness power spectra of the two surfaces are shown in Fig. \ref{Powerspectrum_Roughness3micron_vs_Roughness1micron-Reference.eps}. The surfaces are
self-affine fractal for the wavenumber $q>3\times 10^5 \ {\rm m}^{-1}$, with the Hurst exponent $H=0.8$. We have calculated the pressure and shear stress flow factors 
using the theory of Ref. \cite{PS1}, and the results are shown in Fig. \ref{FluidFlowFactors_Roughness3micron_vs_Roughness1micron-Reference.eps}. The figure shows
$\phi_{\rm p}$ and $\phi_{\rm s}$ as a function of the average interface separation $\bar u$ normalized by the rms height $h_{\rm rms}$. Note that as
a function of $\bar u/h_{\rm rms}$ both power spectra gives the same flow factors, as indeed expected because the two power spectra differ only by a
prefactor. 

\begin{figure}
\includegraphics[width=1.0\columnwidth]{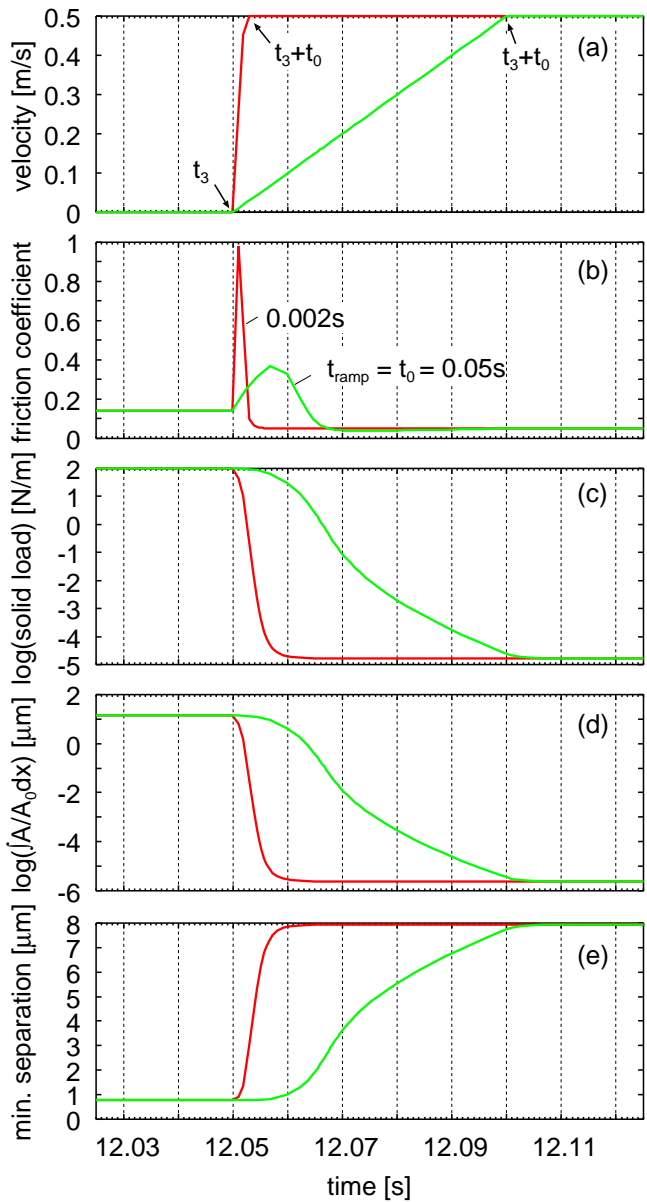}
\caption{\label{RampProfiles_Roughness1micron.eps}
The (a) velocity, (b) friction coefficient, (c) 
load for solid-solid contact, (d) relative contact area, and (e) minimum separation, as functions of time for ramp 
velocity profile with different ramping rates: red curves for ramp time of $t_{\rm ramp}=t_0=0.002 \ {\rm s}$ and green curves for ramp time of $0.05 \ {\rm s}$.
For the normal load $100 \ {\rm N/m}$, rubber cylinder radius $R=4 \ {\rm mm}$, surface roughness amplitude $h_{\rm rms}=1 \ {\rm \mu m}$, elastic modulus $E=3 \ {\rm MPa}$ and lubricant viscosity of $0.1 \ {\rm Pa s}$.
For the ramp profile (b) in Fig. \ref{sinus.eps} with (for red curve): $t_0=0.002 \ {\rm s}$, $t_1-t_0 =2  \ {\rm s}$, $t_2-t_1=0$ and $t_3-t_2=10  \ {\rm s}$. For the green curves we used the same time interval
except $t_0=0.05 \ {\rm s}$. The red curve is shifted by $0.048 \ {\rm s}$ to larger times in order for the start of ramping to occur at the same time point in the figure.
}
\end{figure}

\vskip 0.1cm
{\bf 3.1 Steady sliding}

We first present results for the Stribeck curve, i.e., the friction coefficient as a function of the sliding speed. In the calculations below we
always used a Newtonian liquid with the viscosity $\eta=0.1 \ {\rm MPa}$ 
as is typical for a hydrocarbon lubrication oil. The frictional shear stress $\tau_1$ acting in the
area of real contact is assumed to be $\tau_1 = 1 \ {\rm MPa}$.
Fig. \ref{StribeckCurve_Roughness3micron_vs_Roughness1micron-Reference.eps}
shows (a) the friction coefficient, (b) actual area of contact, (c) minimum separation 
and (d) the solid load, as functions of velocity for the surfaces with $h_{\rm rms}=3 \ {\rm \mu m}$ (red curves) and
$h_{\rm rms}=1 \ {\rm \mu m}$ (green curves). In Fig. \ref{StribeckCurve_Roughness3micron_vs_Roughness1micron-Reference.eps}(a) 
the dashed lines show the contribution of solid-solid contact to the friction coefficient, and the full lines the total friction coefficients.
Note that the surface with the larger surface roughness exhibits a peak in the friction coefficient before entering into the boundary lubrication region.
This peak is due to the friction factor $\phi_{\rm f}$ as can be understood as follows. 
When the surface roughness amplitude increases, the velocity where the first asperity contact occurs will 
shift to higher sliding speeds. As shown in Fig. \ref{StribeckCurve_Roughness3micron_vs_Roughness1micron-Reference.eps}(b) the first contact occur roughly at one decade
higher velocity for the  $h_{\rm rms}=3 \ {\rm \mu m}$ surface as compared to the $h_{\rm rms}=1 \ {\rm \mu m}$ surface.  
When the area of real contact increases, the area where the surface separation is very small (say of order nm) will also increase. In this area the frictional shear stress is
given by $\eta v /u$ where $u$ is the surface separation.

Thus, when the surface roughness is high enough there will be an important contribution to the
friction force from shearing surface regions where the surfaces are separated by a very small distance, say a few nm or so. 
This is manifested in the theory above by $\phi_{\rm f} >> 1$ (see Fig. \ref{explain.eps} and Sec. 2.2).
This contribution will be reduced at smaller sliding speeds because the 
shear rate is proportional to $v$. It will also decrease at higher speeds because then there will be no region
where the surface separation is very small. Hence, if the surface roughness is big enough, we expect a peak in the friction coefficient close to (but below) the velocity
where the first contact occur between the surfaces. We note that this result is for Newtonian fluids. If the fluid exhibit shear thinning the effect we discussed
may be absent. We also note that a peak in the friction coefficient has been observed for sliding friction experiments with glycerol as the lubricant \cite{Scaraggi}.
The effect was only observed when the surface roughness was large enough, in agreement with the results obtained here. 

\begin{figure}
\includegraphics[width=1.0\columnwidth]{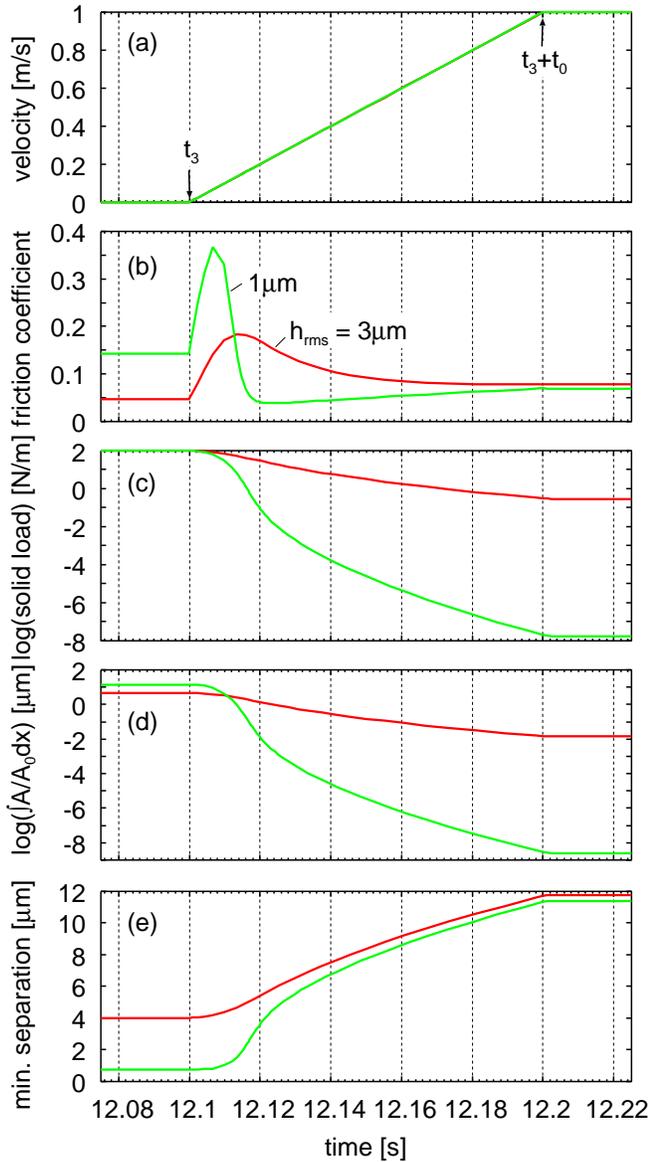}
\caption{\label{RampProfile_Roughness3micron_vs_Roughness1micron-Reference.eps}
The (a) velocity, (b) friction coefficient, (c) load for solid-solid contact, (d) relative contact area, and (e) minimum separation, as functions of time for ramp 
velocity regime for surfaces with different roughness: red curves for $h_{\rm rms}=3 \ {\rm  \mu m}$, green curves for $h_{\rm rms}=1 \ {\rm \mu m}$.  
Ramp time $0.1 \ {\rm s}$, maximum velocity $1 \ {\rm m/s}$, normal load of $100 \ {\rm N/m}$, 
rubber cylinder radius $R=4 \ {\rm mm}$, elastic modulus $E=3 \ {\rm MPa}$ and lubricant viscosity of $0.1 \ {\rm Pa s}$.
For the ramp profile (b) in Fig. \ref{sinus.eps} with (for both red and green curves): $t_0=0.1 \ {\rm s}$, $t_1-t_0 =2  \ {\rm s}$, $t_2-t_1=0$ and $t_3-t_2=10  \ {\rm s}$.
}
\end{figure}

\vskip 0.1cm
{\bf 3.2 Linear multi-ramp motion}

Let us now consider non-stationary sliding. We assume first a multi-ramp case 
where the driving velocity depends on time as indicated in Fig. \ref{sinus.eps}(b). 
The most interesting results are for a time period around the time $t=t_3$ of the start of the second ramping of the velocity.
In Fig. \ref{RampProfiles_Roughness1micron.eps} we show (a) the velocity, (b) friction coefficient, (c) 
load for solid-solid contact, (d) relative contact area, and (e) the minimum separation, 
as functions of time for two ramp 
velocity profiles with different ramping rates: red curves for ramp time of $t_{\rm ramp}=t_0=0.002 \ {\rm s}$ and green curves for ramp time of $0.05 \ {\rm s}$.
For the normal load $100 \ {\rm N/m}$, rubber 
cylinder radius $R=4 \ {\rm mm}$, surface roughness amplitude $h_{\rm rms}=1 \ {\rm \mu m}$, elastic modulus $E=3 \ {\rm MPa}$ and lubricant viscosity of $0.1 \ {\rm Pa s}$.
Note the large peak in the friction for the faster ramping. This is again due to the shear stress term $\eta v /u$. Thus, at the start of ramping the 
average surface separation is small and the area of real contact large. Thus there will be relatively large regions between the surfaces where the surface separation is
very small (nanometers) and shearing the thin fluid film in these regions will give an important contribution to the friction force, which is the origin of the large 
peak in the friction force observed in Fig. \ref{RampProfiles_Roughness1micron.eps}(b). We will see in the following (see Sec. 4) that
the existence of this friction peak in independent experimental results. Furthermore, it follows that the breakloose friction force observed in
many experiments, e.g., for syringes, may have an important contribution from shearing the non-contact, lubricant filled regions with small surface separation, i.e.,
the breakloose friction for is not solely due to shearing the area of real contact as assumed in some studies of the breakloose friction force\cite{Tabor,Squeeze}.

Fig. \ref{RampProfile_Roughness3micron_vs_Roughness1micron-Reference.eps}
shows the same as in Fig. \ref{RampProfiles_Roughness1micron.eps}, but now for ramping the velocity linear to $v_0=1 \ {\rm m/s}$ during $0.1 \ {\rm s}$.
Results are shown for the two different surfaces with  $h_{\rm rms}=3 \ {\rm  \mu m}$ (red curves) and $h_{\rm rms}=1 \ {\rm \mu m}$ (green curves).  
Note that for the smoother surface the friction peak during ramping is higher and more narrow (as a function of time) than for the rougher surface. 
The peak is due to the shearing of the fluid film, and since the smoother surface, before the start of the velocity ramp, has larger surface area with small
surface separation $u$ than for the rougher surface, the term $\eta v/u$, when integrated over the surface area, will be larger for the smoother surface.
The more narrow width of the friction peak result from the fact that as the speed increases the fluid pressure buildup will separate the surfaces,
and complete separation occur faster for the smoother surface while the rougher surface still will have some surface regions with small separation $u$ at relatively high sliding speed, which
will result in a contribution from shearing the surface regions with small separation extending to higher sliding speeds for the more rough surface.

\begin{figure}
\includegraphics[width=1.0\columnwidth]{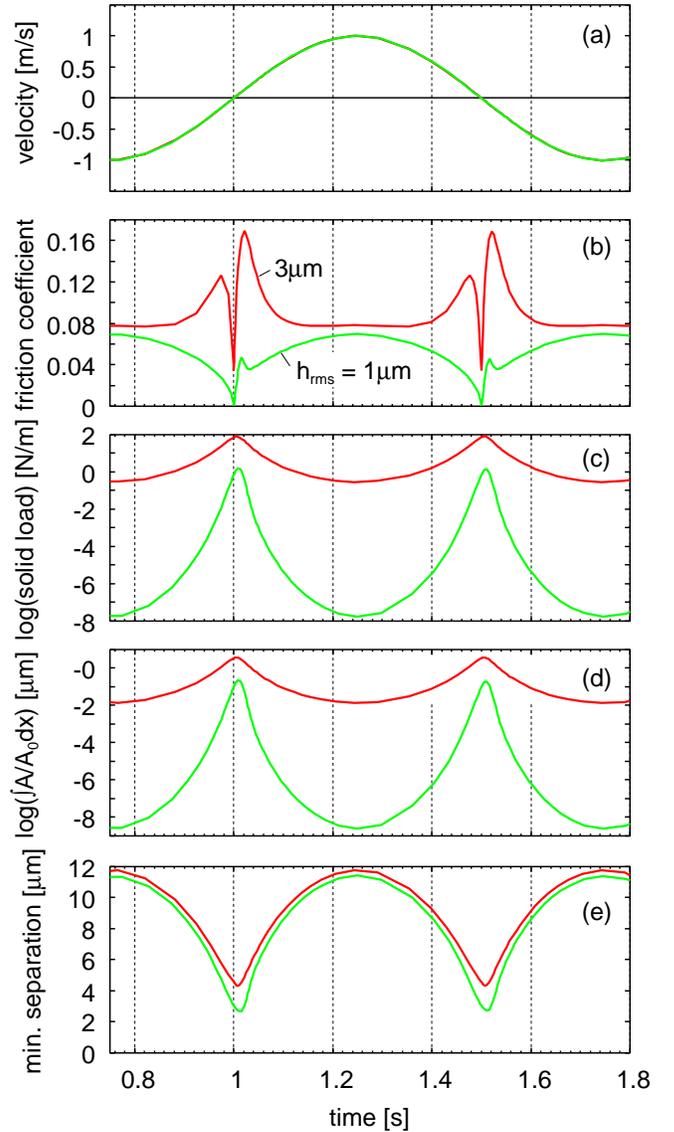}
\caption{\label{SinusProfile_Roughness3micron_vs_Roughness1micron-Reference.eps}
The (a) velocity, (b) friction coefficient, (c) load for solid-solid contact, (d) relative contact area, and (e) minimum separation, as functions of time for sinusoidal reciprocating motion
with different roughness: red curves for $h_{\rm rms}=3 \ {\rm \mu m}$ and green curves for $h_{\rm rms}=1 \ {\rm \mu m}$. 
For the reciprocating frequency $1 \ {\rm Hz}$, velocity amplitude of $1 \ {\rm m/s}$, normal load of $100 \ {\rm N/m}$, 
rubber cylinder radius $R=4 \ {\rm mm}$, elastic modulus $E=3 \ {\rm MPa}$ and lubricant viscosity of $0.1 \ {\rm Pa s}$.
}
\end{figure}

\vskip 0.1cm
{\bf 3.3 Sinus sliding motion}

Fig. \ref{SinusProfile_Roughness3micron_vs_Roughness1micron-Reference.eps}
shows the same results as in Fig. \ref{RampProfile_Roughness3micron_vs_Roughness1micron-Reference.eps} 
but now for sinusoidal reciprocating motion (as in Fig. \ref{sinus.eps}(a)).

\begin{figure}[tbp]
\includegraphics[width=1.0\columnwidth]{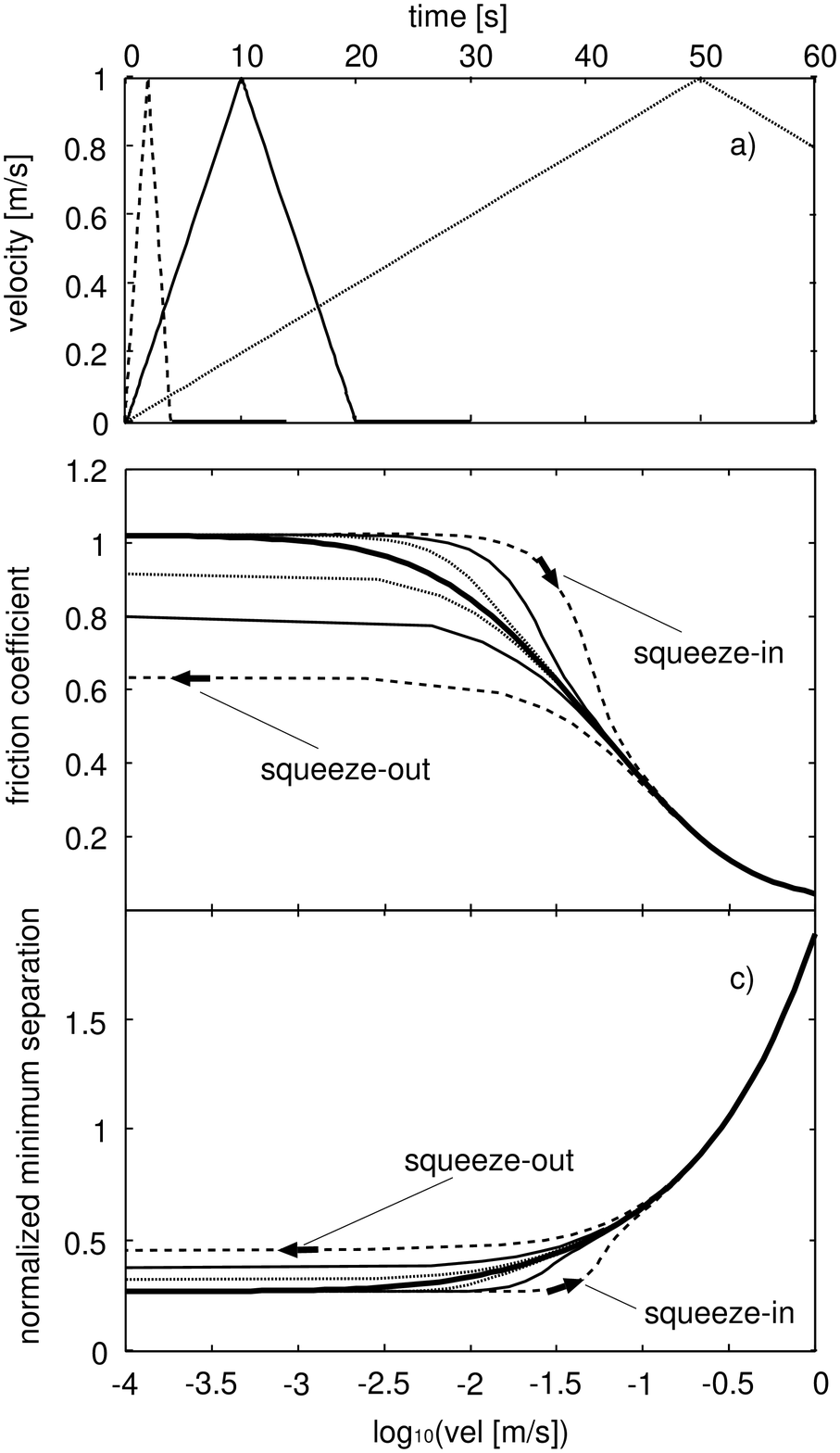}
\caption{\label{combined.eps} Friction (b) and minimum locally-averaged interface gap (c) as a function of the sliding speed in
log scale, for the sliding kinematics reported in (a). The sliding motion (a) is obtained by
constant acceleration from 0 up to 1 {\rm m/s}, and then constant deceleration up to stop. Four accelerations $a$ values are adopted,
with the steady sliding case corresponding to $a\rightarrow 0$ (solid thick line in (b) and (c)).
The arrows in (b) and (c) show the time direction. For the normal load $777.5 \ {\rm N/m}$, rubber 
cylinder radius $R=2.5 \ {\rm mm}$, isotropic surface roughness with $h_{\rm rms}=2.4 \ {\rm \mu m}$,
low frequency cut-off $q_0=0.311\ 10^3\ \mathrm{m}$, high frequency cut-off $q_1=5.9\ 10^7\ \mathrm{m}$,
roll-off $q_\mathrm{r}=q_0$ and fractal dimension 2.2. Elastic modulus $E=3 \ {\rm MPa}$ (Poisson ratio $\nu=0.5$)
and Newtonian lubricant with viscosity $0.1 \ {\rm Pa s}$. $\sigma_\mathrm{f}=10\ MPa$.
}
\end{figure}

Note that there is an asymmetry in the friction coefficient around the time-points where the velocity vanishes. This is due to the time dependency of the
fluid squeeze-out: during the fast motion ($v\approx 1 \ {\rm m/s}$) the surface separation is relatively large (about $12 \ {\rm \mu m}$ in both cases).
\begin{figure}[tbp]
\includegraphics[width=1.0\columnwidth]{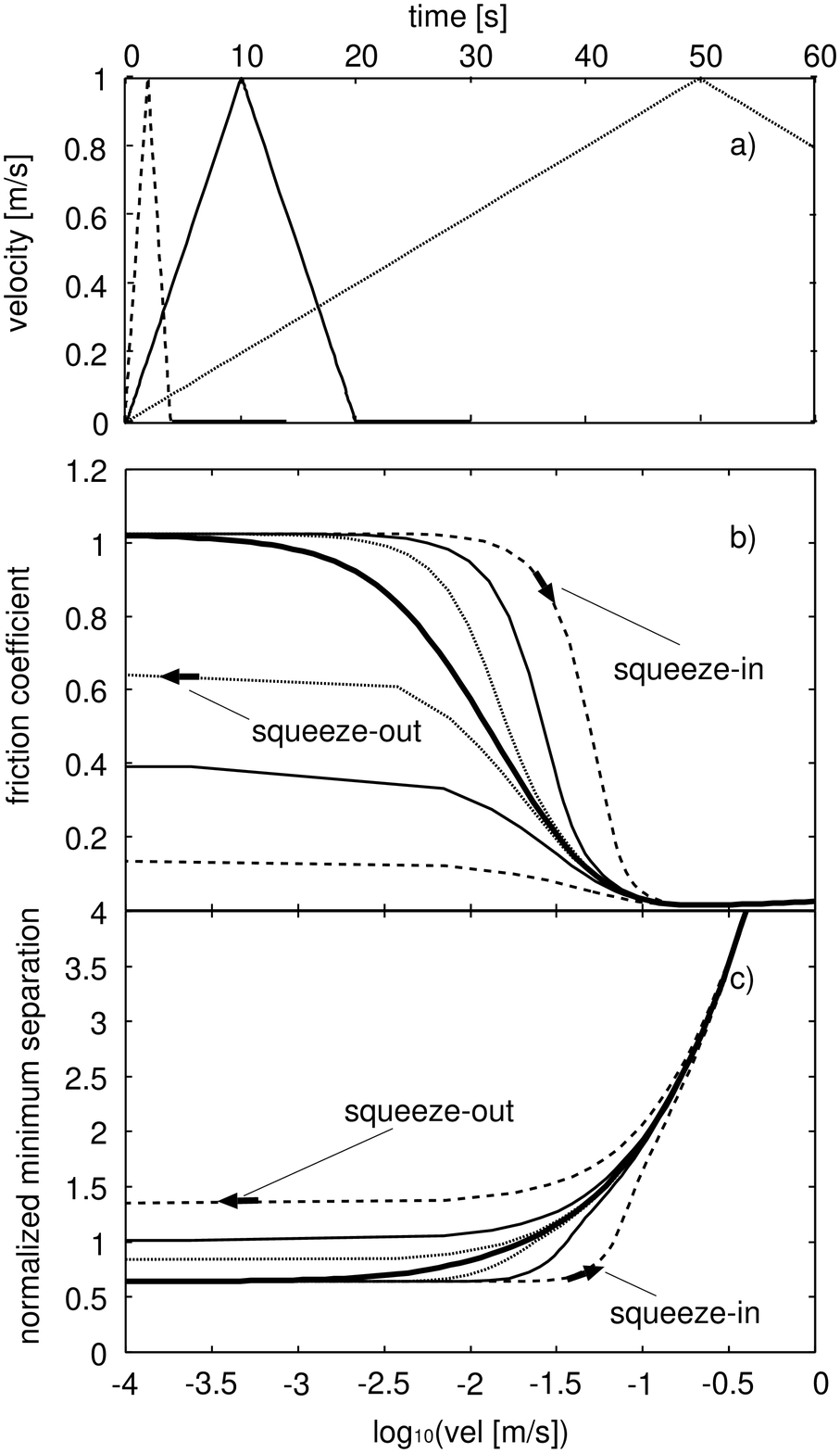}
\caption{\label{combined.2.eps} Friction (b) and minimum locally-averaged interface gap (c) as a function of the sliding speed in
log scale, for the sliding kinematics reported in (a). In particular, the sliding motion (a) is obtained by
constant acceleration from 0 up to 1 {\rm m/s}, and then constant deceleration up to stop. Four accelerations $a$ values are adopted,
with the steady sliding case corresponding to $a\rightarrow 0$ (solid thick line in (b) and (c)).
The arrows in (b) and (c) show the time direction. For the same parameters of Fig. \ref{combined.eps} but for
a cylinder radius $R=2.5 \ {\rm cm}$.
}
\end{figure}
\begin{figure*}[tbp]
\includegraphics[width=1.8\columnwidth]{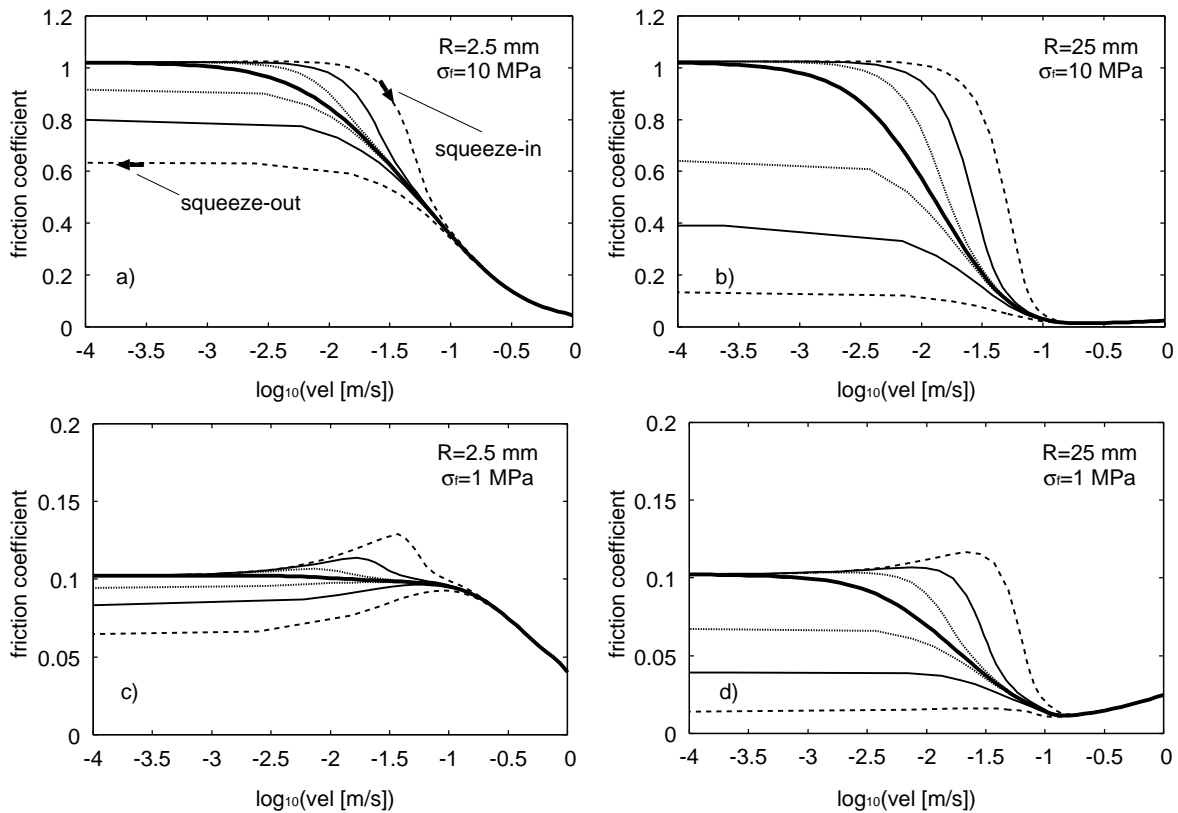}
\caption{\label{combined.4.eps} Friction as a function of the sliding speed in
log scale, for the same parameters of Fig. \ref{combined.eps} but for
(a) a cylinder radius $R=2.5 \ {\rm mm}$ and $\sigma_\mathrm{f}=10 \ {\rm MPa}$, (b) a cylinder radius $R=2.5 \ {\rm cm}$ and $\sigma_\mathrm{f}=10 \ {\rm MPa}$,
(c) a cylinder radius $R=2.5 \ {\rm mm}$ and $\sigma_\mathrm{f}=1 \ {\rm MPa}$ and (d) a cylinder radius $R=2.5 \ {\rm cm}$ and $\sigma_\mathrm{f}=1 \ {\rm MPa}$.
}
\end{figure*}
As the velocity decreases towards zero the surface separation decreases, but this decrease continue for a short time interval even during the increase in the
velocity beyond the time-points where $v=0$. Thus, the minimum surface separation, and the local maximum in the friction coefficient, occur slightly after the
time-points where the velocity vanish.
This type of asymmetry in the time-dependent friction coefficient has been observed experimentally\cite{Pegg} (see Sec. 4). 
Note also that the friction peaks are much higher for the larger surface roughness case. 
This is because the surface separation in the low-velocity range
is still rather large, and only for the large-roughness case is the area of contact, and the region where the surface separation 
is very small (where hence the shear stress $\eta v/u$ is large), high.
This explains why for the small roughness surface a (small) friction peak is observed 
only after the velocity has changes sign, while for the large roughness case, (large) friction peaks are observed on both sides of the $v=0$ time-points. 


\vskip 0.1cm
{\bf 3.4 The role of squeeze-out and squeeze-in on friction}

During accelerated sliding, squeeze-in and squeeze-out can be defined as the normal (to the contact) interface motion caused by,
respectively, a positive and negative rate of variation of the average interface separation. Thus,
during squeeze-in a fluid flow, driven by pressure gradient, will occur toward the contact in order to replenish the interface
up to the new interface separation value, the latter dictated by the accelerated motion. Since a finite time, proportional to the fluid viscosity,
is required for the squeeze-in process to occur, a larger friction than for the steady-sliding
is expected during this replenishment motion, determined by the smaller average contact gap (thus, larger fluid and
solid contact friction). Similar but inverse considerations apply for the squeeze-out process.

We observe that the breakloose friction, i.e. the friction measured during the start-up of a machine element, will thus endure
more than expected (i.e. on the basis of the rate of variation of sliding or rolling speeds) because of this
finite time associated to the squeeze-in phenomenon. It would be therefore very interesting to quantify,
even for a particular contact case, the fluid squeeze effects on the friction. Thus in Fig. \ref{combined.eps}
we show calculation results, in term of friction (b) and minimum locally-averaged interface gap (c) as a function of the sliding speed, for
the sliding kinematics reported in Fig. \ref{combined.eps}(a). In particular, the sliding motion is obtained by
constant acceleration from 0 up to 1 {\rm m/s}, and then constant deceleration up to stop. Four accelerations $a=0$, $0.02$, $0.1$ and $0.5 \ {\rm m/s^2}$ values are adopted,
with the steady sliding case corresponding to $a\rightarrow 0$ (solid thick line in Fig. \ref{combined.eps}(b) and (c)), which we refer to as the Stribeck curve.
The arrows in Fig. \ref{combined.eps}(b) and (c) show the time direction.

We note first that all the friction curves
lie on the Stribeck curve during the first acceleration instants (note: since time scales linearly with velocity, $v=at$,
the log scale in Fig. \ref{combined.eps}(b) and (c) during acceleration corresponds to a log scale in times, too).
This is due to the initial condition assumed, for
all the simulations, given by a rough-Hertzian initial condition (i.e. without oil entrapment at start-up, see e.g. the for
hard interactions \cite{jmps}). At increasing
sliding speeds, however, the squeeze-in process occurs leading to an enlargement of the breakloose friction plateau (say, the boundary regime)
toward increasing velocities (see dashed curve in Fig. \ref{combined.eps}(b)). This effect is more severe for larger accelerations, and it involves
a contact range belonging to the mixed lubrication regime.
This extended frictional plateau corresponds to an extended plateau in the minimum film thickness value, as shown in Fig. \ref{combined.eps}(c). 
By further increasing the sliding speed, all the minimum gap and friction curves converge to the master steady-sliding curve.

At decreasing sliding speeds, instead, the squeeze-out process occurs leading to an extended plateau in the minimum separation. Interestingly,
the minimum gap is almost doubled for the fastest motion. As a consequence, a plateau is obtained in the friction curves, with strongly reduced
friction coefficient, as clearly shown in Fig. \ref{combined.eps}(b). Furthermore, we observe that the initial and final contact conditions differs because
of squeeze dynamics. The latter involves complex percolation mechanisms at the interface \cite{Mueser}, and in particular under large
normal pressures, the solid contact area can percolate in an annular region close to the Hertzian contact circle, leading to
a mechanically stable lubricant entrapment. In such a case, only very slow (long time scales) inter-diffusion processes, where the trapped islands of pressurized fluid diffuse
into the rubber (and from there perhaps to the external environment) can lead the
lubricant to escape from the trapping.

Similar considerations apply to Fig. \ref{combined.2.eps}, where we have simulated for the same parameters of Fig. \ref{combined.eps} but
for a cylinder radius $R=2.5 \ {\rm cm}$. As expected, the larger radius (thus, the larger Hertzian area) increases the strength of the squeeze out dynamics effects,
leading to a larger extension of the frictional plateau during squeeze-in, and to a smaller friction value during squeeze-out.
Finally, in Fig. \ref{combined.4.eps} we show the effect of the shear stress acting in the true contact area $\sigma_\mathrm{f}$ and of the cylinder radius on the squeeze dynamics and observable friction. We note that at reduced values of $\sigma_\mathrm{f}$, during the start of ramping (squeeze-in motion), a peak occurs in the friction curves (Figs. \ref{combined.4.eps}c and \ref{combined.4.eps}d) instead of the plateau discussed before (Figs. \ref{combined.4.eps}a and \ref{combined.4.eps}b). This is due to the reduction of the adhesive
contribution to dissipation (occurring in the true contact areas) which allows the shearing action, occurring in the nanometers-separated fluid-filled areas, to 
increase its weight in the total friction, similarly to the large 
peak in the friction observed in Fig. \ref{RampProfiles_Roughness1micron.eps}(b).

\begin{figure}[tbp]
\begin{center}
\includegraphics[width=1.0\columnwidth]{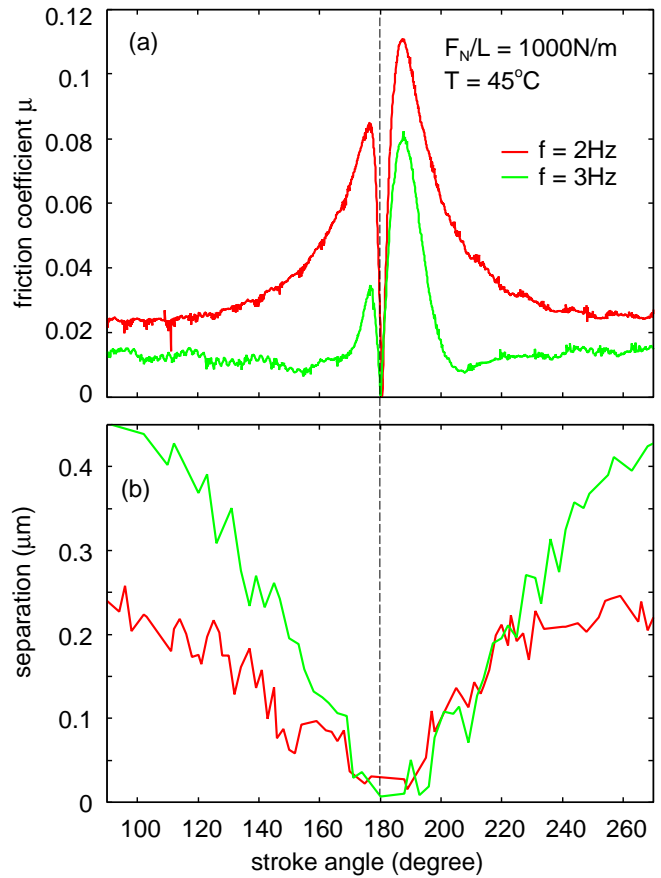}
\end{center}
\caption{
The (a) friction coefficient and (b) minimum separation for a steel cylinder (radius $R=4 \ {\rm cm}$)
in contact with  a fused silica specimen with a flat surface lubricated by an oil with the viscosity $0.062 \ {\rm Pas}$
at $T=45^\circ {\rm C}$ (temperature at the measurement). The silica disk is oscillating at the frequency $f=2 \ {\rm Hz}$ (red curve)
or $3 \ {\rm Hz}$ (green curve). The stroke length is $d=2.86 \ {\rm cm}$ and the normal load per unit length
$F_{\rm N}/L = 1000 \ {\rm N/m}$. Based on experimental data from Ref. \cite{Pegg}. }
\label{1angle.2mu.and.d.2Hz.3Hz.10N.eps}
\end{figure}

\begin{figure}[tbp]
\begin{center}
\includegraphics[width=1.0\columnwidth]{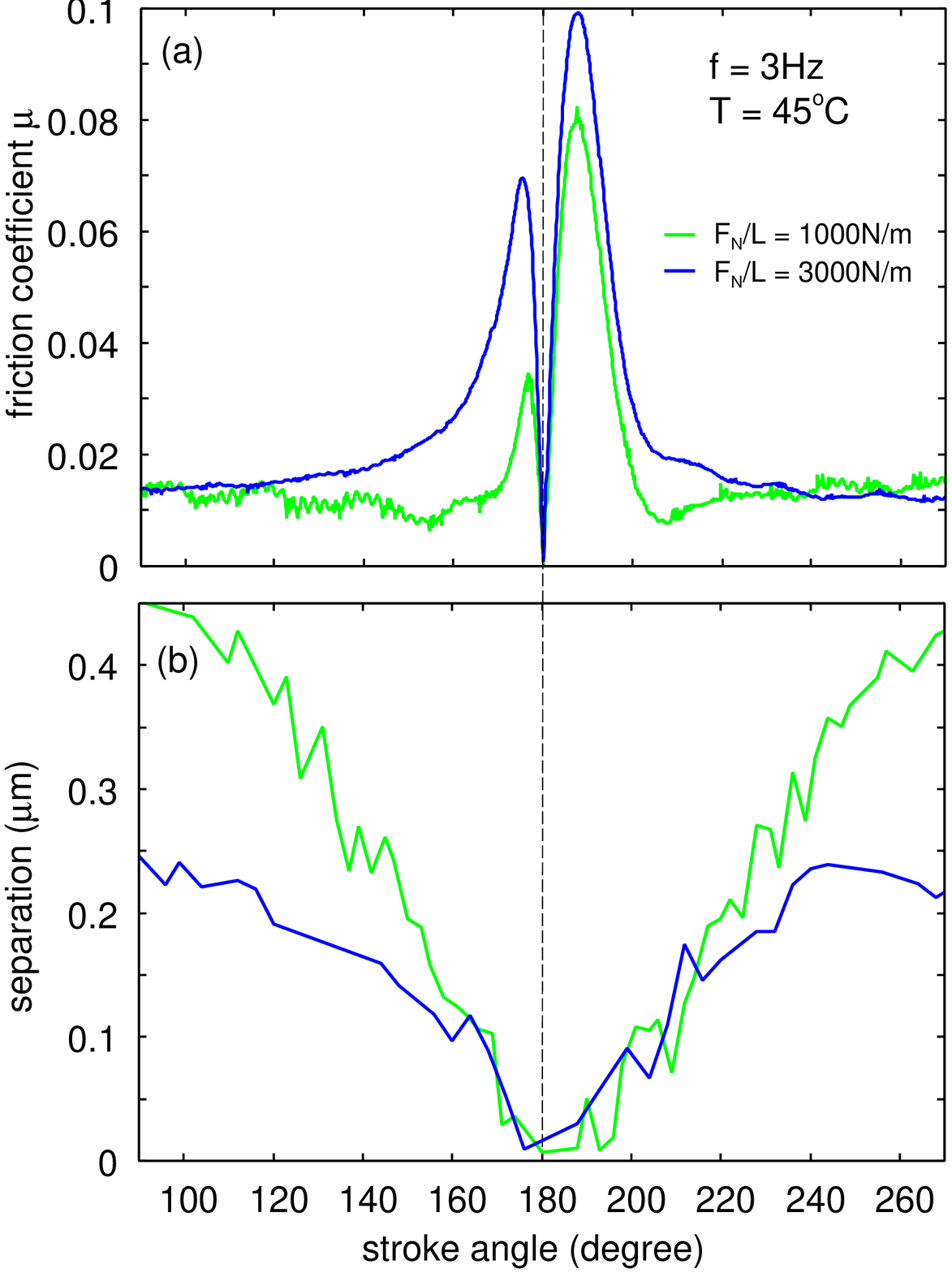}
\end{center}
\caption{
The (a) friction coefficient and (b) minimum separation for a steel cylinder (radius $R=4 \ {\rm cm}$)
in contact with  a fused silica specimen with a flat surface lubricated by an oil with the viscosity $0.062 \ {\rm Pas}$
at $T=45^\circ {\rm C}$ (temperature at the measurement).
The normal load per unit length
$F_{\rm N}/L = 1000 \ {\rm N/m}$ (red curve) and $3000 \ {\rm N/m}$ (green curve). 
The stroke length is $d=2.86 \ {\rm cm}$ and the silica disk is oscillating at the frequency $f=3 \ {\rm Hz}$.
Based on experimental data from Ref. \cite{Pegg}.}
\label{1angle.2mu.and.d.10N.30N.3Hz.eps}
\end{figure}

\begin{figure}[tbp]
\begin{center}
\includegraphics[width=1.0\columnwidth]{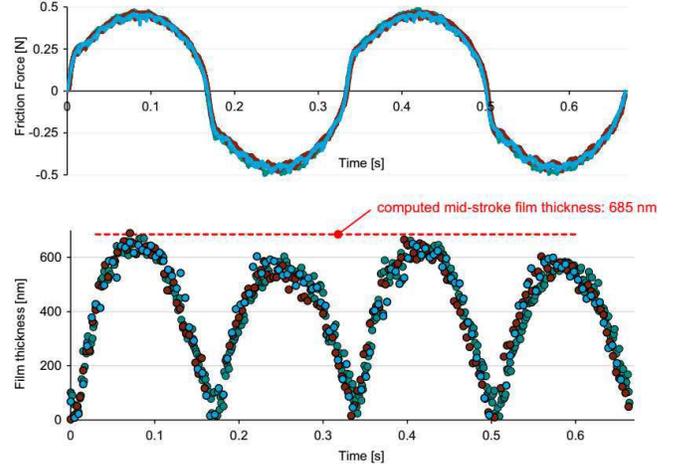}
\end{center}
\caption{
Friction coefficient (top) and minimum separation (bottom) for a steel cylinder (radius $R=4 \ {\rm cm}$)
in contact with  a fused silica specimen with a flat surface lubricated by an oil with the viscosity $0.27 \ {\rm Pas}$
at $T=15^\circ {\rm C}$ (temperature at the measurement).
The normal load per unit length
$F_{\rm N}/L = 3000 \ {\rm N/m}$. 
The stroke length is $d=2.86 \ {\rm cm}$ and the silica disk is oscillating at the frequency $f=3 \ {\rm Hz}$.
Adapted from Vladescu et al. \cite{Pegg}.}
\label{uk.eps}
\end{figure}

\begin{figure}[tbp]
\begin{center}
\includegraphics[width=1.0\columnwidth]{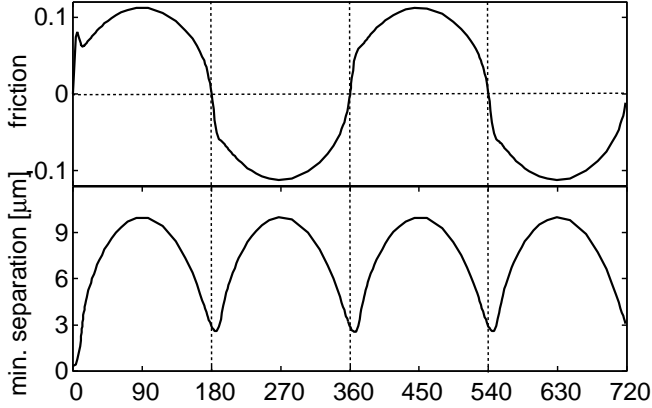}
\end{center}
\caption{
Friction coefficient (top) and minimum separation (bottom) for an elastic rough cylinder (radius $R=1 \ {\rm mm}$, $E_\mathrm{r}=3.95 \ {\rm MPa}$)
in alternating sinus sliding contact with a rigid flat surface, lubricated by a Newtonian oil with viscosity $0.1 \ {\rm Pas}$.
The normal load per unit length
$F_{\rm N}/L = 117 \ {\rm N/m}$, whereas the shear stress acting in the true contact areas is assumed $\sigma_\mathrm{f}=1 \ {\rm MPa}.$
The stroke length is $d=0.1 \ {\rm m}$ and the stroke time is $T=0.1 \ {\rm s}$. The cylinder is covered
by an isotropic roughness characterized by $q_0=1\times 10^4\ {\rm m^{-1}}$, $q_\mathrm{r}=3\times 10^5\ {\rm m^{-1}}$,
$q_1=3\times 10^9\ {\rm m^{-1}}$, $h_{\rm rms}=1\ {\rm \mu m}$ and fractal dimension $D_{\rm f}=2$.}
\label{fzj.eps}
\end{figure}

\begin{figure}[tbp]
\begin{center}
\includegraphics[width=1.0\columnwidth]{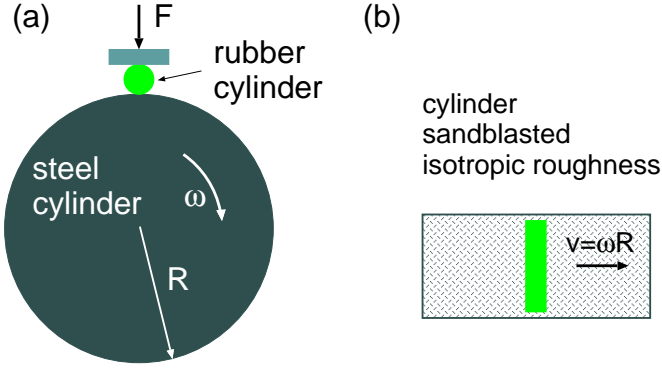}
\end{center}
\caption{(a) Schematic picture of the experimental friction tester. The rubber cylinder is pushed
with a dead weight towards the rotating steel cylinder. (b) The steel cylinder has surface roughness
prepared by sandblasting (bottom). The latter results in surface roughness with isotropic
statistical properties.}
\label{1}
\end{figure}

\begin{figure}[tbp]
\begin{center}
\includegraphics[width=1.0\columnwidth]{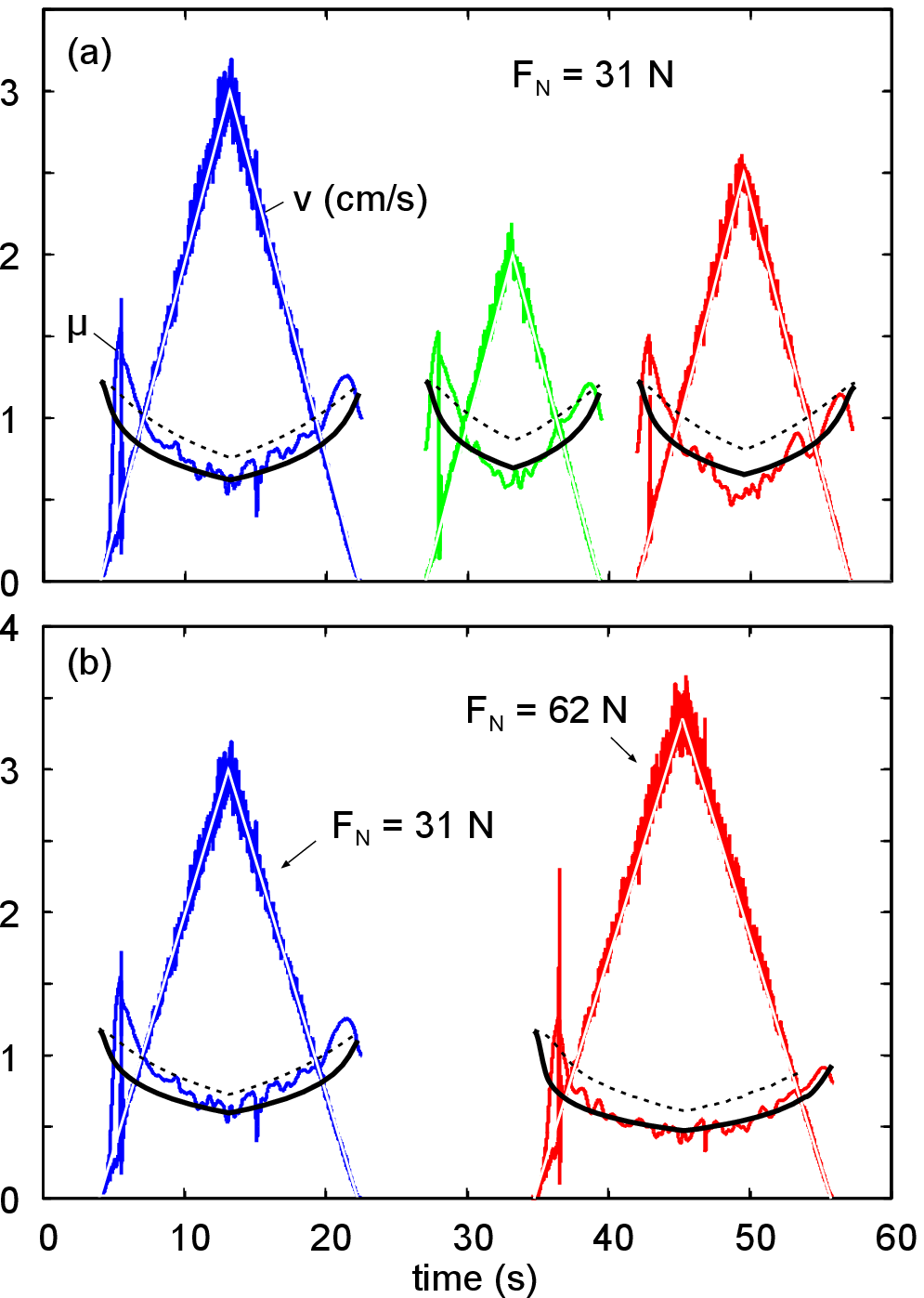}
\end{center}
\caption{The friction coefficient, and the rotation velocity of the steel cylinder,
as a function of time. The velocity of the steel cylinder first increase linearly with time and then decrease linearly with
time with the same absolute value for the acceleration. In (a) we show results for three cases where the maximum velocity
differ but the load (or normal force) is constant $F_{\rm N}=31 \ {\rm N}$. In (b) we show results for two different normal load,
$F_{\rm N}=31 \ {\rm N}$ and $F_{\rm N}=62 \ {\rm N}$. $\sigma_\mathrm{f}=11.5\ {\rm MPa}$.
}
\label{1time.2mu.and.v.31N.and.62N.eps}
\end{figure}

\vskip 0.3cm
{\bf 4 Experimental results}

\vskip 0.1cm
{\bf 4.1 Sinus sliding motion}

The results presented in Sec. 3.3 are in qualitative agreement with experimental observation.
Thus, Vladescu et al. \cite{Pegg} have performed experiments where a steel
cylinder with the radius of curvature $R=4 \ {\rm cm}$ was slid in reciprocating motion (stroke length $2.86 \ {\rm cm}$,
frequency $f=1$, $2$ or $3 \ {\rm Hz}$) on
a flat fused silica glass surface. The steel surface has the rms-roughness $18 \ {\rm nm}$
when measured over a $431  {\rm \mu m} \times 575 {\rm \mu m}$ surface area. The interface was lubricated
with an oil with the viscosity  $0.062 \ {\rm Pas}$
at $T=45^\circ {\rm C}$ (the temperature during the measurement).

Fig. \ref{1angle.2mu.and.d.2Hz.3Hz.10N.eps} and \ref{1angle.2mu.and.d.10N.30N.3Hz.eps} shows
(a) the friction coefficient and (b) the minimum separation between the steel cylinder and the 
glass surface. Fig.  \ref{1angle.2mu.and.d.2Hz.3Hz.10N.eps} shows results when 
the silica disk is oscillating at the frequency $f=2 \ {\rm Hz}$ (red curve) or $3 \ {\rm Hz}$ (green curve),
with the normal load per unit length $F_{\rm N}/L = 1000 \ {\rm N/m}$.
Fig. \ref{1angle.2mu.and.d.10N.30N.3Hz.eps} shows results for the normal load per unit length
$F_{\rm N}/L = 1000 \ {\rm N/m}$ (red curve) and $3000 \ {\rm N/m}$ (green curve), with the 
silica disk oscillating at the frequency $f=3 \ {\rm Hz}$.

Note that the results in Fig. \ref{1angle.2mu.and.d.2Hz.3Hz.10N.eps} and \ref{1angle.2mu.and.d.10N.30N.3Hz.eps} 
are qualitatively identical to what we observe in our calculations, 
see Fig. \ref{SinusProfile_Roughness3micron_vs_Roughness1micron-Reference.eps} (b) and (e).
In particular, the friction peak just after reversal of the sliding direction is larger than the friction peak
just before reversal of the sliding direction. This is also found in the theory and is due to the longer squeeze-out
time in the former case. Note also that the minimum in the surface separation as a function of the stroke angle is displaced
slightly to the right of the turn-around angle $\alpha = 180^\circ$. This is again due to the longer squeeze-out time
to the right of the turn-around angle. As expected, increasing the frequency from $f=2$ to $3 \ {\rm Hz}$ result in
lower friction and larger surface separation due to the build-up of a higher hydrodynamic pressure in the lubricant film as a result of the
increase in the sliding speed. Similarly, increasing the load from $F_{\rm N}/L = 1000$ to $3000 \ {\rm N/m}$
reduces the oil film thickness and increases the friction.

At the moment we cannot replicate numerically the results reported by Vladescu et al. \cite{Pegg}; indeed,
under a load of $F_{\rm N}/L = 1 \ {\rm kN/m}$, and for the given materials properties
($E_1=210\ \mathrm{GPa}$ and $\nu_1=0.29$ for steel, $E_2=73\ \mathrm{GPa}$ and $\nu_2=0.17$ for the fused silica), the Hertzian semicontact
length is about 0.95 {\rm mm}, whereas at $F_{\rm N}/L = 3 \ {\rm kN/m}$ one finds about 1.6 {\rm mm}. Considering that the ring
is 2 {\rm mm} thick, this means that the interaction is not occurring under a Hertzian-like condition (i.e. the contact is
extended to the ring edges) and thus the shape of the ring edges will strongly determine the hydrodynamic lift. However,
similarly to what reported before, the main dynamical features of the lubricated contact are in very good agreement with the theory.
This is confirmed also in the comparison between Fig. \ref{uk.eps} (adapted from Vladescu et al. \cite{Pegg}) and our results
Fig. \ref{fzj.eps}. In particular, on the top and bottom figure we show, respectively, the friction and the minimum separation
for a cylinder in sliding reciprocating motion. It is interesting to observe that the experimental friction curves exhibit a localised friction
spike during motion reversal (due to a squeeze-out prolonged over the beginning of the accelerated motion), which is
in qualitative agreement with the theory.

\vskip 0.1cm
{\bf 4.2 Linear multi-ramp motion}

In order to experimentally investigate the lubricated line contact of a
generic hydraulic seal, a test rig has been designed and set up at the
Institute for Fluid Power Drives and Controls (IFAS).
A steel cylinder with radius $R=20 \ {\rm cm}$ is rotated at varying angular
speed $\omega$, and squeezed in contact with a $L=4 \ {\rm cm}$ long Nitrile Butadiene Rubber (NBR) cylinder (segment of
an o-ring) with diameter $D = 0.5 \ {\rm cm}$. A normal force 
$F_{\rm N}$ is applied to the contact, see see Fig. \ref{1}(a).
The rubber cylinder is fixed in space while the steel cylinder can be let to rotate either with a constant speed or else 
in accelerated motion. 
The rubber cylinder is assumed to have a perfectly smooth surface while the steel surface is sandblasted
with the rms-roughness $2 \ {\rm \mu m}$.
The lubricant fluid is a standard hydraulic oil with the room-temperature viscosity $\eta \approx 0.1 \ {\rm Pa s}$.

Fig. \ref{1time.2mu.and.v.31N.and.62N.eps} shows the friction coefficient, and the peripheral velocity of the steel cylinder,
as a function of time. The velocity of the steel cylinder first increases linearly with time and then decreases linearly with
time with the same absolute value as for the accelerated stage. In (a) we show results for three cases where the maximum velocity
differ but the normal force is constant $F_{\rm N}=31 \ {\rm N}$. In (b) we show results for two different normal load,
$F_{\rm N}=31 \ {\rm N}$ and $F_{\rm N}=62 \ {\rm N}$. Points (with colors) are experimental results, whilst the dashed line is from the theory
presented in Sec. 2 (roughness on the fixed cylinder), whereas the solid line is again from theory but applied to the case corresponding to the experimental setup
(roughness on the moving cylinder, see the complementary theory in the companion paper \cite{new}). We observe that the agreement is very good, 
unless for the very beginning of the ramp motion where the lateral deformation dynamics of the instrumented measurement arm plays a role
in the formation of the breaklose friction value (so called elastic sliding). However, we also note, interestingly, that 
including in the calculations the roughness on the top fixed (dashed line) solid instead of on the bottom sliding (solid line) solid leads to qualitatively different
friction results, suggesting the importance of the correct evaluation of flow and friction factors in soft contacts.

\vskip 0.3cm
{\bf 5 Summary and conclusion}

We have extended the theory developed in Ref. \cite{PS0,PS1,PS2,SP0} in order to
study non-stationary (transient) elastohydrodynamic problems including surface roughness,
non-Newtonian liquid lubrication, and arbitrary accelerated motion.
We have presented several illustrations for an elastic cylinder with randomly rough surface sliding on a perfectly
flat and rigid substrate lubricated by a Newtonian fluid (no shear thinning). We considered both reciprocal
motion ($v=v_0 {\rm sin}(\omega t)$) and linear multi-ramp motion. The calculated results were compared to
experimental data and very good qualitative agreement was obtained. We plan to perform 
sliding friction experiments of the type described above 
on surfaces with known (measured) surface roughness power spectra to compare quantitatively
to the theory predictions. We will report on these results elsewhere.

\vskip 0.2cm
{\it Acknowledgments:}
We thank S-C Vladescu and T. Reddyhoff (Ref. \cite{Pegg}) for supplying the numerical data used for Fig. \ref{1angle.2mu.and.d.2Hz.3Hz.10N.eps} and \ref{1angle.2mu.and.d.10N.30N.3Hz.eps}.
This work was performed within a Reinhart-Koselleck project funded by the Deutsche Forschungsgemeinschaft (DFG).
We would like to thank DFG for the project support under the reference German Research Foundation DFG-Grant: MU 1225/36-1.
The research work was also supported by the DFG-grant: PE 807/10-1 and DFG-grant No. HE 4466/34-1.
MS acknowledges FZJ for the support and the kind hospitality
received during his visit to the PGI-1. Finally, MS also acknowledges COST Action MP1303 for
grant STSM-MP1303-171016-080763.


\begin{thebibliography}{99}

\bibitem{Persson0}
B.N.J. Persson, {\it Sliding Friction: Physical Principles and Applications}, Springer, Heidelberg (2000).

\bibitem{Meyer}
E. Gnecco and E. Meyer, {\it Elements of Friction Theory and Nanotribology}, Cambridge University Press (2015).

\bibitem{Greg}
A.C. Dunn, J.A. Tichy, J.M. Uruena and W.G. Sawyer, Tribology International {\bf 63}, 45 (2013). 

\bibitem{Mueser}
W.B. Dapp, A. Lucke, B.N.J. Persson and M.H. M\"user,  Phys. Rev.Lett. {\bf  108}, 244301 (2012).

\bibitem{Dowson}
D. Dowson and G.R. Higginson, J. Mech. Egrs. Sci. {\bf 1}, 6 (1959).

\bibitem{elasto}
R. Gohar, {\it Elastohydrodynamics}, second edition, World Scientific Publishing, Singapore (2001).

\bibitem{PS1}
B.N.J. Persson and M. Scaraggi, Eur. Phys. J. E{\bf 34}, 113 (2011).

\bibitem{PS0}
B.N.J. Persson and M. Scaraggi, J. Phys.: Condens. Matter {\bf 21}, 185002 (2009)

\bibitem{PS2}
B.N.J. Persson, J. Phys.: Condens. Matter {\bf 22}, 265004 (2010)

\bibitem{PS3}
M. Scaraggi, G. Carbone, B.N.J. Persson and D. Dini
Soft Matter {\bf 7}, 10395 (2011).

\bibitem{SP0}
M. Scaraggi, B.N.J. Persson,
Tribology letters {\bf 47}, 409 (2012).

\bibitem{SP1}
B. Lorenz, B.N.J. Persson
The European Physical Journal E {\bf 32}, 281 (2010).

\bibitem{PS.intsep}
B.N.J. Persson, Phys. Rev.Lett. {\bf 99}, 125502 (2007).

\bibitem{Scaraggi}
M. Scaraggi and G. Carbone and D. Dini, Trib. Lett. {\bf 43}(2), 169-174 (2011). 

\bibitem{Tabor}
A.D. Roberts and D. Tabor, Proc. R. Soc. Lond. A {\bf 325}, 323 (1971).

\bibitem{Squeeze}
B. Lorenz, B.A. Krick, N Rodriguez, W.G. Sawyer, P. Mangiagalli and B.N.J Persson,
J. Phys.: Condens. Matter {\bf 25}, 445013 (2013).

\bibitem{Pegg}
S-C Vladescu, S. Medina, A.V. Olver, I.G. Pegg and T. Reddyhoff, 
Tribology International {\bf 98}, 317 (2016).

\bibitem{jmps}
M. Scaraggi and G. Carbone, JMPS {\bf 58}(9), 1361-1373 (2010).
 
\bibitem{new}
M. Scaraggi, J. Angerhausen, L. Dorogin, H. Murrenhoff and B.N.J. Persson, submitted (2017).

\end{thebibliography}
\end{document}